\documentclass[a4paper]{article}
\usepackage[margin=25mm]{geometry}
\usepackage{amsfonts}
\usepackage{amssymb}
\usepackage{graphicx}
\usepackage{svg}
\usepackage{amsmath}
\usepackage{caption}
\usepackage{etoolbox}
\usepackage{bm}
\usepackage[labelfont={small,bf}]{caption}
\usepackage[font={small}]{caption}
\usepackage{subcaption}
\usepackage{hyperref}
\usepackage{authblk}
\usepackage[numbers,sort&compress]{natbib}

\title{The unreasonable likelihood of being \\[0.5em]
\large Origin of life, terraforming, and AI\\ \ \\}

\author{Robert G. Endres}

\affil{Department of Life Sciences, Imperial College, London SW7 2AZ, United Kingdom}

\begin{document}
\maketitle

\begin{abstract}
The origin of life on Earth via the spontaneous emergence of a protocell prior to Darwinian evolution remains a fundamental open question in physics and chemistry. Here, we develop a conceptual framework based on information theory and algorithmic complexity. Using estimates grounded in modern computational models, we evaluate the difficulty of assembling structured biological information under plausible prebiotic conditions. Our results highlight the formidable entropic and informational barriers to forming a viable protocell within the available window of Earth’s early history. While the idea of Earth being terraformed by advanced extraterrestrials might violate Occam’s razor from within mainstream science, directed panspermia—originally proposed by Francis Crick and Leslie Orgel—remains a speculative but logically open alternative. Ultimately, uncovering physical principles for life’s spontaneous emergence remains a grand challenge for biological physics.
\end{abstract}

\section*{Beginnings}\label{sec:into}

“All cells come from cells” \cite{Virchow1855} leads us into a classic chicken-and-egg dilemma: where did the first cell come from? Either it came from somewhere else—conveniently outsourcing the mystery—or it emerged from the laws of physics and chemistry on a young, chaotic, and geologically active Earth. Once a minimal protocell or replicator emerged, Darwinian evolution could take over, merely selecting and diversifying its way toward greater complexity, biodiversity, and eventually, the modern biosphere, from gold fish to particle accelerators and social media. One thing this onset of a discussion illustrates - how little we know about our origin, and the cosmos in general.

As for the “somewhere else” hypothesis, science fiction has long entertained terraforming scenarios. In 2001: A Space Odyssey, an inscrutable black monolith gives early hominins a mysterious push toward cognition—and tool use \cite{Kubrick1968}. In Star Trek, entire planetary ecosystems are engineered by advanced civilizations or errant AI, often with remarkable speed \cite{Roddenberry1966}. The X-Files proposed that alien viruses might have seeded Earth, albeit oddly fixated on rural America \cite{Carter1993}. In Prometheus, a lone extraterrestrial—credited only as “the Engineer”—dissolves himself in a waterfall, presumably catalyzing Earth’s biochemistry, though the narrative coherence dissolves nearly as fast \cite{Scott2012}. The mother of all terraforming stories is, however, the Book of Genesis, written more than 2,500 years ago. It declares that life was created in six days \cite{GenesisBible} — a claim that has profoundly shaped human thought, even if not amenable to scientific testing.

\begin{figure}[t]
\centering
\includegraphics[scale=0.3]{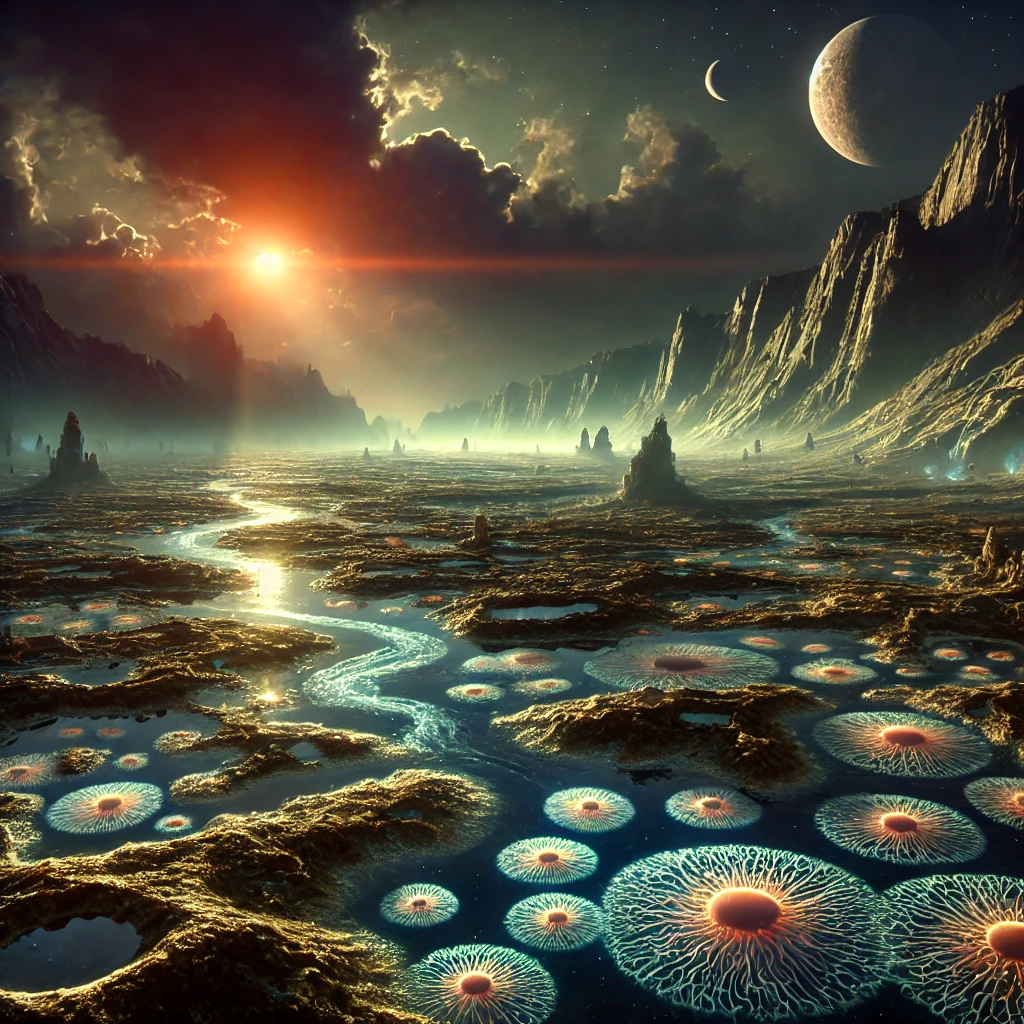}
\caption{{\bf Fantasy sci-fi imagery of terraforming.} Chatgpt4.0's hallucination of early Earth with seeded biomaterial, jump starting Darwinian evolution.}
\label{fig:fig1}
\end{figure}

One step closer to science (though not necessarily to certainty), Francis Crick—the co-discoverer of DNA’s structure—and Leslie Orgel proposed in 1973 the idea of directed panspermia \cite{Crick1973}. In their scenario, an advanced extraterrestrial civilization, facing extinction or perhaps scientific curiosity, dispatches microbial “starter kits” to habitable planets like ours (see Fig. \ref{fig:fig1} for an illustration). While Crick and Orgel attempted to formulate this idea more like a testable hypothesis, it deftly relocates the explanatory burden to someone else’s biochemistry.

Meanwhile, modern Bayesian analyses of life’s emergence struggle with deeply unconstrained data (or lack thereof): a sample size of one, and priors shaped not only by empirical observation but also by immense subjectivity \cite{Spiegel2012,Kipping2020}. There’s also the problem of timing—life appeared early on Earth, yet the intelligent observer required to marvel this fact arrived very much later, introducing a strong observational bias. In this work, we revisit the origin of life through the lens of modern cell models and AI-based tools to better appreciate the informational and physical complexity required for protocell formation. We begin with general considerations and propose a qualitative theoretical framework. Finally, we follow Feynman's mantra that "doubt and uncertainty are not to be feared but welcomed" \cite{Feynman1955}, and attempt to cautiously conclude something.

\section*{Geological constraints can be soft}

\noindent When did life emerge, and what are the soft and hard constraints? There were two massive impacts with global sterilizing effects: Earth was first struck by a Mars-sized body, named ``Theia,'' about $4.51$\,Gy ago, leading to the formation of the Moon \cite{Hartmann1975}. This was followed by another sterilizing impact, dubbed ``Moneta,'' approximately $40$\,My later \cite{Benner2020},  {and the Late Heavy Bombardment of many smaller impacts between $\sim 4.1$ to $3.8$\,Gy, which, in contrast to the earlier giant impacts, are not believed to have been globally sterilizing}. Mineralogical evidence from zircons indicates that both an atmosphere and liquid water were already present on Earth’s surface around $4.404$\,Gy ago \cite{Wilde2001}. The oldest putative evidence for life comes from $^{13}$C-depleted carbon inclusions in $4.1$\,Gy-old zircon deposits \cite{Bell2015}. The oldest direct and undisputed evidence is found in microfossils preserved in $3.465$\,Gy-old rocks in Western Australia \cite{Schopf2006,Schopf2007,Schopf2018}.

 {Around the same time, the planet’s core was already generating a global magnetic field: palaeomagnetic signatures imply a functioning geodynamo by $\geq 3.5$\,Gy as evidenced by permanently magnetized basalt and dacite rocks, and possibly as early as $4.2$\,Gy, though such early zircon records remain debated \cite{Tarduno2015HadeanGeodynamo,Olson2013CoreParadox}. This magnetic field likely played an important role in enabling life to establish and persist on the surface of the Earth by shielding it from the damaging solar wind.}

Recently, a study combining molecular clock methodology, horizontal gene-transfer-aware phylogenetic reconciliation, and existing biogeochemical models addressed the nature of the last universal common ancestor (LUCA), estimating its age, gene content, metabolism, and ecological context \cite{Moody2024}. Surprisingly, LUCA is estimated to have lived around $\sim 4.2$ Gy ago, based on calibration with fossil and isotope records. More perplexing still, LUCA appears to have been an anaerobic acetogen, metabolically similar to modern prokaryotes, and already equipped with ATP synthesis, a TCA cycle, early immune capabilities including CRISPR-Cas effector proteins, and embedded within a microbial ecosystem. A recent consensus across eight genomic and proteomic studies broadly supports this view \cite{Crapitto2022}. This suggests that a protocell—presumably far simpler—must have emerged even earlier, perhaps relying on RNA or other primitive replicator systems.

Only in the last blink of planetary time — some 100,000 years ago — did anatomically modern humans appear, wielding curiosity and stunning technological inventions. Although
the controlled use of fire by earlier hominins dates back about 1 Myr \cite{wrangham2009catching}, in geological terms intelligence has arrived extraordinarily late. 
Yet anthropic considerations must be taken into account: without human observers, it would not have been possible  to ponder this spectacle.
Intriguingly, this late arrival may be further constrained from the other end of the timeline, as Earth’s biosphere  to support human life is projected to collapse in roughly 100 Myr due to impending “CO$_2$ starvation”
(even way before the sun becomes a Red Giant and engulfs the Earth) \cite{Lovelock1982Nature,Caldeira1992}. As solar luminosity increases, silicate weathering accelerates, drawing down atmospheric CO$_2$. Below $\sim 150$ ppm, modern plant photosynthesis ceases, triggering ecological collapse \footnote{ {Photosynthetic pathways differ in their tolerance to low CO$_2$ concentrations. While the biosphere with some cacti and microbes can survive much longer,  $\approx 1$ GYrs, this is about $C_4$ photosynthesis which can work for even lower CO${}_2$ levels ($<10$ ppm CO${}_2$). However, most trees, crops, and grass are C${}_3$ photosynthesis based, and hence end much earlier (at about 150 ppm CO${}_2$). Hence, all animal (and human) life will likely end in about 100 MYrs.}}. Hence, if intelligent life requires too much time to emerge, many planets—especially around hotter, short-lived stars—may simply run out of time, biasing observations toward an unusually early start.

 {These considerations highlight that the geological and environmental constraints discussed above are best understood as soft, probabilistic limits rather than hard impossibilities. Such probabilistic limits can produce truly fascinating, even optimistic outlooks:} a modified Drake equation—updated with exoplanet data—places a firm lower bound on the likelihood that at least one other technological species has arisen at some point in the observable universe. If the probability that a habitable-zone planet spawns a technological species is $\geq 10^{-24}$ — about the odds of winning the lottery a modest $8$ times in a row — then humanity is almost certainly not alone \cite{Frank2016}. So statistically speaking, we’re probably not the only ones wondering whether we’re alone.

\section*{Physical principles of life's emergence - anyone?}

\noindent Since Erwin Schrödinger’s landmark 1944 essay \textit{What is Life?}, physicists have speculated whether the emergence of life might require physical principles beyond those already known \cite{Schrodinger1944}. Schrödinger speculated—presciently, though vaguely—about undiscovered laws governing order and organization in living systems, prompting generations of researchers to explore the physics of self-organization. Since then, the field has progressed dramatically. Here is a short list without any claim of completeness:

Alan Turing’s work on reaction–diffusion systems introduced a mathematical framework for how spatial patterns, such as those observed in morphogenesis, can spontaneously arise from simple rules \cite{Turing1952}. The Landauer–Bennett principle revealed a deep connection between information, entropy, and the thermodynamic cost of computation—key concepts for understanding living systems as information-processing entities \cite{LeffRex2003,Lloyd2006}. Ilya Prigogine’s theory of dissipative structures demonstrated how local order can emerge in open systems that export entropy \cite{Prigogine1984}. Stuart Kauffman’s concept of autocatalytic sets offered a compelling model for how chemical networks might self-amplify and sustain themselves \cite{Kauffman1993}. The latter two demonstrate nicely that complex systems have new emergent collective features beyond the sum of their parts, not predictable from first principles \cite{Anderson1972}.

More recently, in work bordering the popular science literature, Jeremy England has proposed that life-like order could emerge generically through driven dissipation under physical selection-like conditions \cite{England2020}, while Sara Walker has emphasized the centrality of information flow and causal structure in defining what it means to be alive \cite{Walker2022}.  {In this spirit, together with Lee Cronin, the emphasis on the history of paths for making complex molecules led to assembly theory (revisited later) \cite{Sharma2023Assembly}. Common to all the above is the argument that to understand life’s origin, we must look beyond molecular inventories and instead focus on the dynamics of information and energy, and thus causation as organizational principles in physical systems.
Cast into the language of non-equilibrium physics, the likelihood of reaching a given nonequilibrium state need not be set by the thermodynamic properties of the state itself, but by the entropy produced along the transition path. As shown for multistable non-equilibrium systems,  switching with high dissipation can reliably populate even statistically rare states \cite{cook2020thermodynamics}. The origin of life may therefore be reducible to a sequence of exergonic transitions, with path thermodynamics rather than state stability governing the assembly process.}

Broadly speaking, origin-of-life relevant research in physics (often theoretical) falls within the rich and rapidly evolving domains of active matter, stochastic thermodynamics, and nonequilibrium phase transitions. These fields have significantly expanded our understanding of how systems far from equilibrium can maintain structure and perform work. Yet despite this progress, a fully satisfying mechanism for the spontaneous emergence of a protocell from disordered chemistry remains elusive.  {While biological processes obey physical laws, the laws themselves appear agnostic to the emergence of life, given our present understanding.}

\section*{Dogmas and artificial intelligence}

 {\noindent Beyond formal physical laws, the societal context constrains what is considered worthwhile to work on, as scientific dogmas define domains of accepted knowledge, enabling structured peer review and funding decisions.} Yet they can also inhibit new discoveries \cite{Kuhn1962}. For example, the phrase “nothing in biology makes sense except in the light of evolution” \cite{Dobzhansky1973}, though foundational in laboratory practice and bioinformatics, may inadvertently create boundaries that hinder unconventional thinking about life’s origins. Similarly, the dictum “all cells come from cells” \cite{Virchow1855} tends to obscure the deeper question of how the first cell came to be.  {Even within the origin-of-life community, research remains notably siloed: RNA-world chemists, mineral-surface and hydrothermal-vent advocates, astrobiologists, exoplanet scientists, and theoretical physicists working on stochastic thermodynamics, active matter, or phase transitions often pursue largely independent agendas. This fragmentation is widely recognised in the field, with repeated calls for integrating geochemistry, planetary science, and prebiotic chemistry \cite{Sutherland2017EndBeginning,CleavesBada2012AlternativeNucleicAcids,Hazen2007emergence}.}

 {Against this backdrop of conceptual and disciplinary constraints, one thing is certain:} with the rise of powerful AI tools and mechanistic models, we now have entirely new ways of exploring biological complexity. Tools like AlphaFold for protein folding \cite{Jumper2021} and comprehensive whole-cell models \cite{Crick1973_ProjectK,Karr2012,AhnHorst2022} allow us to estimate the information content of life using Kolmogorov (algorithmic) complexity. This measures the shortest description required to reproduce a system—the more structured or regular a system, the more it can be compressed and the easier it can be made \cite{LiVitanyi2008}. Here, by combining such complexity estimates with basic rate-distortion theory \cite{Shannon1959,CoverThomas2006}, we aim to shed new light on the plausibility of life’s rapid emergence.

\section*{Theoretical framework}\label{sec:results}

\noindent We approach the problem of life's origin by estimating how plausible it is for chemical processes to spontaneously assemble a sufficiently complex protocell within the available time on early Earth. We assume this window to be about $500$ Myr—interpolated between “optimistic” ($304$ Myr) and “conservative” ($939$ Myr) estimates \cite{Kipping2020}. We apply a heuristic version of Shannon's rate-distortion theory \cite{Shannon1959,CoverThomas2006}, which quantifies the minimal information rate $R(D)$ needed to construct a functioning system, given a tolerated distortion level $D$. Here, $D$ characterizes how much loss (error or degradation) can be tolerated in forming a viable protocell. The distortion rate depends on the entropy of the chaotic prebiotic environment, denoted $H_\text{prebiotic}$, which represents the information content of the chemical soup in a pond or hydrothermal vent.

\begin{figure}[t]
\centering
\includegraphics[scale=0.55]{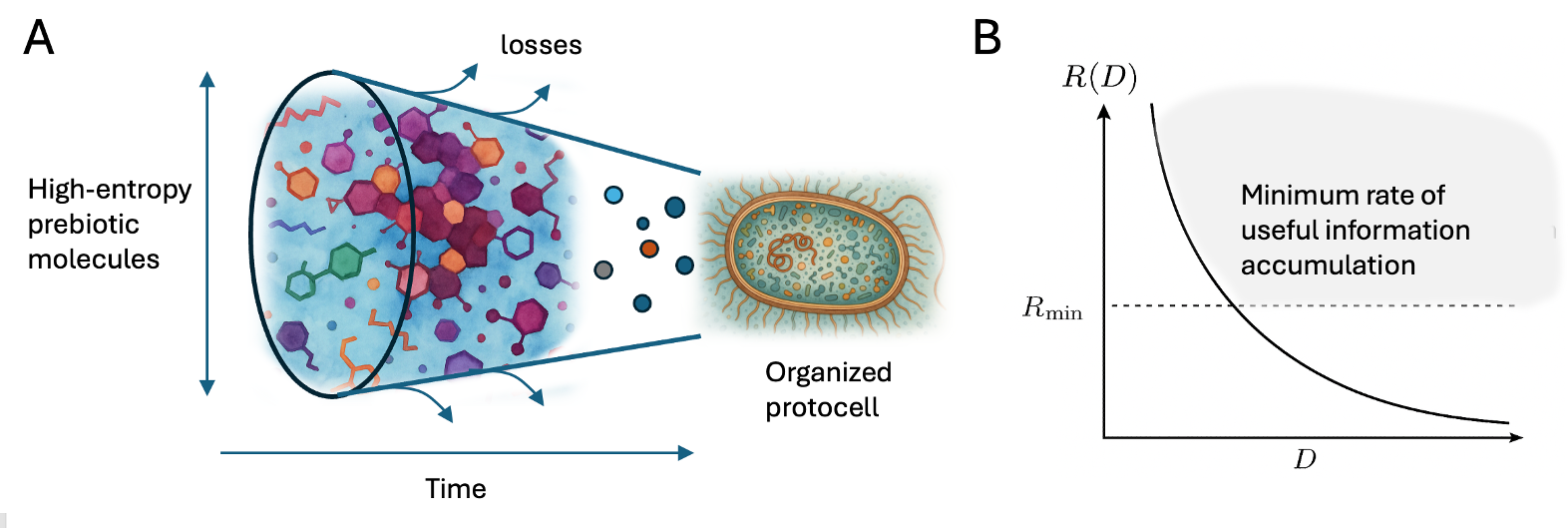}
\caption{{\bf Mathematical framework based on rate-distortion theory.} (A) Funnel of collapsing entropy for the emergence of a protocell.  {While this is easily drawn, the challenge is to do this thermodynamically consistently.} (B) Rate-distortion function from information theory, comparing $R(D)\sim H_\text{prebiotic}/D$ as a function of $D$ with $R_\text{min}$ from observational constraints (see the main text for explanations).}
\label{fig:fig2}
\end{figure}

To succeed, the rate $R(D)$ must be at least as large as the macroscopic information rate required to assemble a protocell of complexity $I_\text{protocell}$ within the time available, $T_\text{available} = 500$ Myr (see Fig. \ref{fig:fig2}). We begin by writing in mathematical form:
\begin{equation}
   R(D)\approx \underbrace{\eta \frac{H_\text{prebiotic}}{D}}_\text{entropy collapse} 
   \geq 
   \underbrace{R_\text{min}=\frac{I_\text{protocell}}{T_\text{available}}}_\text{rise in complexity}
   \label{eq:framework}
\end{equation}
 
\noindent In this simplified formulation, we approximate $R(D)$, typically convex and decreasing, by the ratio of prebiotic entropy, $H_\text{prebiotic}$, and molecular persistence time, $D$,  {where $\eta<\!\!<1$} is an efficiency prefactor, introduced here as a compact way to account for microscopic inefficiencies  {(sometimes colloquially referred to as a ``fudge factor''). The factor $\eta$, as harmless as it looks, accounts for the fact that only very few molecular encounters actually result in usable contributions. Molecules may be structurally misaligned, chemically unstable, or fail to integrate into a coherent assembly due to the absence of repair mechanisms or spatial constraints in the local environment. Determining this factor is really at the heart of the missing physical theory of protocell formation.}

 {As one concrete illustration of the complexities subsumed into $\eta$, we consider the so-called homochirality problem, i.e.\ the difficulty of assembling a protocell in a racemic environment. Most biological molecules — amino acids, sugars, nucleotides — exist in two mirror-image forms (L and D), identical in energy but incompatible in function, yet all amino acids in biology are L and all ribose sugars in biology are D, requiring demixing and purification,  effectively at a global scale. And even though the chiral symmetry breaking problem looks solved since the 1950s in a closed, well-mixed, kinetically idealized environment \cite{Frank1953}, it is hard to underestimate the problem of the constant influx of wrong chirality molecules from the outside, terminating biochemical pathways and thus jamming protocell assembly \cite{Soai1995, Ozturk2023,Laurent2024}.}

Taken together, all quantities in Eq.~\ref{eq:framework}—except perhaps $T_\text{available}$—are highly uncertain. Nevertheless, following this line of reasoning, the spontaneous origin of life remains plausible if the left-hand side exceeds the right-hand side. If not, we may be missing some crucial ingredient—such as additional evolutionary pathways, unknown physical principles, or external interventions like Crick and Orgel’s directed panspermia (or, more colorfully, aliens terraforming the planet). For the sake of argument, we set aside the "cheap" statistical fallback that rare events inevitably occur somewhere in a vast universe.

\section*{Entropy of prebiotic soup and tolerated error rate of assembly}

In the absence of a time machine, we estimate conditions on Earth roughly 4 Gy ago. We begin with the left-hand side of Eq.~\ref{eq:framework}, aiming to estimate the entropy of the chemical soup, $H_\text{prebiotic}$, and the tolerated error rate $D$—which characterizes how long complex organic molecules persist before degrading. Importantly, the entropy $H_\text{prebiotic}$ pertains not to an entire ocean or vent system, but to the volume accessible to a nascent protocell. While the volume of a single protocell may be around 1 femtolitre ($10^{-15}$ l), relevant molecules could diffuse in from a larger surrounding droplet. We seek order-of-magnitude estimates informed by chemical experiments simulating early Earth conditions, meteorite data, and planetary spectroscopy.

In the Miller–Urey experiments, spark discharges in a reducing gas mixture produced a variety of amino acids and other small organic molecules \cite{Miller1953}. During the Cassini–Huygens mission, analysis of Titan’s atmosphere revealed that complex photochemistry can generate thousands of distinct organic species, including nitriles, hydrocarbons, and possible prebiotic precursors \cite{Horst2017}. Similarly, carbonaceous chondrite meteorites (e.g., Murchison) contain more than 60 amino acids, as well as nucleobases, polyols, carboxylic acids, and related compounds \cite{Sephton2002}. This is consistent with samples returned from asteroid Bennu by the OSIRIS-REx mission, which revealed a rich inventory of organic compounds—including those preserved in salt-rich brine inclusions in phyllosilicate minerals, dating back to $\sim 4.5$ billion years \cite{McCoy2025}.

Combinatorial chemistry across diverse geochemical settings could plausibly yield thousands of unique organic molecules, suggesting a chemical library on the order of $10^5$–$10^6$ small molecules in the prebiotic environment. Assuming an average information content of $\sim 10$ bits per molecule—representing functional or structural specificity—this allows for roughly $10^3$ distinguishable states per molecule (e.g., different side chains, folding propensities, or reactivities). 
 {This estimate aligns with protein domain annotations and bits-per-residue values from sequence--function maps \cite{Durston2007,AdamiCerf2000,Weiss2000ProteinInfo}, including quantitative analyses of sequence information content in natural proteins \cite{Possenti2018ProteinInfoPartition}. 
Related bounds also arise from canonical chemical representations such as SMARTS and InChI, whose string lengths provide effective upper limits on molecular descriptive complexity \cite{Heller2015}. }
Multiplying, we estimate the prebiotic entropy as $H_{\text{prebiotic}} \sim 10^6$–$10^7$\,bits.

Next, we consider the distortion timescale $D$, interpreted as the average lifetime of a complex organic molecule before degradation by hydrolysis, UV damage, oxidation, or further transformation. This defines the “error rate” or persistence window for useful molecular structures. Under plausible early-Earth conditions, complex organics such as ribose and peptides have half-lives ranging from hours to days. For example, ribose degrades with a half-life of $\sim 3$ hours at pH 7 and 37$^\circ$C \cite{Larralde1995}. Small peptides degrade over hours to days depending on sequence and pH. Nucleobases like adenine are sensitive to solar UV and degrade within days to weeks. Without an ozone layer, degradation could be faster—on the scale of hours to days.
Conservatively, we adopt a mean lifetime of $\sim 1$ day, or about $10^5$ seconds, for moderately stable molecules in a warm, aqueous, anoxic environment exposed to UV. However, in cryogenic or mineral-protected contexts—such as asteroid interiors—organic molecules may persist for billions of years \cite{McCoy2025}. The existence of such intergalactic freezers implies that $D$ could vary over many orders of magnitude depending on local geochemical context.

Combining $H_\text{prebiotic} \sim 10^6$–$10^7$ bits with $D \sim 10^5$ s and assuming optimistically an efficiency $\eta = 1$, we obtain:
\begin{equation}
R(D) \approx \frac{H_\text{prebiotic}}{D} \sim 100\ \text{bit/s}.
\end{equation}
Thus, chemical evolution must extract or preserve about 100 bits per second to successfully cross the informational threshold required for life's emergence.

\vspace{1ex}
For further intuition, here are two aids helping to think about the meaning of this estimate. Consider, if you like, a pool of $10^7$ bits of potentially useful molecules, each lasting about $10^5$ seconds. To prevent this information from vanishing forever, you need to stabilize or incorporate about 100 bits per second. Or more dramatically, picture a ‘melting library’: you stand in a library with 10 million books ($H_\text{prebiotic}$), each self-destructing 24 hours after removal. Your task is to compile a manual for “building life.” To capture information before it disappears, you’d need to scan roughly 100 books per second.

\vspace{1ex}
If the molecular lifetime were much shorter—say $D \sim 1$ s, due to rapid diffusion or lack of protective mechanisms—then the required information rate would increase accordingly. The key idea remains: if $R(D)$ is too low to meet $R_\text{min} = I_\text{protocell} / T_\text{available}$, then abiotic emergence becomes implausible without invoking directed evolution, external contributions (e.g., panspermia), or unknown self-organizing principles. The delivery of relatively stable, complex molecules—by meteorites or interstellar dust—could act as a bootstrap for prebiotic evolution \cite{McCoy2025}. Given the large uncertainties in parameter values, our aim is not to deliver a definitive conclusion but to structure a quantitative discussion.

\section*{Information content of a protocell}

\noindent The right-hand side of Eq.~\ref{eq:framework} requires an estimate of the complexity of a protocell, $I_\text{protocell}$. How can we quantify this complexity, including its structural and dynamic organization? It shouldn't come as a surprise that cells are not simply bags of well-mixed molecules \cite{deDuve1984GuidedTour}, but highly structured systems—down to the folding and interaction of their constituent proteins. Moreover, cellular processes are orchestrated in time. If proteins are the musicians, then signaling, metabolic, and genetic pathways are the conductor and the score — guiding timing, harmony, and transitions.

Nature’s “code” is embedded in physical law and can often be expressed in elegant mathematical equations—sometimes compact enough to fit on a single sheet of paper. Unlike our computational models, Nature does not compute itself; it simply unfolds. To make sense of this complexity, we turn to symbolic and algorithmic representations, such as equations or computer code. Yet this translation from physical causality to epistemic representation introduces ambiguity: we cannot neatly separate what is essential from what is merely descriptive. As a result, estimates of Kolmogorov complexity are inherently fuzzy. From below, complexity is bounded by simple, universal laws; from above, it is constrained by the size of compressed code needed to realistically simulate the system. But even this upper bound is imperfect—it includes implementation overhead and excludes unknown biological details. In short, the more we compress reality, the more we risk distorting its true complexity.

\vspace{1ex}
To proceed, we divide the total information content of a protocell, somewhat artificially, into distinct components:
\begin{equation}
I_\text{protocell} = I_g + I_s + I_d + I_a \label{eq:I_protocell}
\end{equation}
where $I_g$ is the genetic information, $I_s$ the structural information (e.g. protein folds, cellular organization), and $I_d$ the dynamic information encoded in biochemical pathways. Each term is estimated via Kolmogorov (algorithmic) complexity, defined as the length of the shortest program that can realistically generate the relevant component, at least conceptually. Hence, Kolmogorov complexity describes the information needed to specify the final structure. 

 {However, this description is not yet sufficient. Even if all components were fully specified, physical realization introduces an additional and independent constraint: the complexity of assembly, which captures not only the number of components but also the order, feasibility, and reliability of the assembly steps themselves. While Kolmogorov complexity quantifies the length of a minimal description, assembly complexity reflects the execution of that description under physical constraints. We therefore introduce an additional contribution, $I_a$, accounting for assembly in Eq.~\ref{eq:I_protocell}. This term emphasizes that random assembly cannot generate a complex cell, because the dominant informational bottleneck lies not in specification but in the assembly pathway. Viable biological organization requires nested, hierarchical construction: elementary components assemble into intermediate modules, which in turn form higher-level structures, culminating in the protocell. Each stage reduces entropy locally, but at the cost of increasing dependence on correct ordering and stabilization. Consequently, early protocells must have been extremely simple, with complexity growing only incrementally, as each new organizational layer required robust scaffolding to stabilize those below.}

\vspace{1ex}
The genomic information is the easiest to estimate—and, perhaps surprisingly, contributes the least. Consider a minimal self-replicating organism such as {\it Mycoplasma genitalium}, which contains roughly 500 genes \cite{Karr2012}, each about 1,000 base pairs long. With two bits per base pair (for A, T, C, G), this yields $I_g \sim 500 \times 1000 \times 2 = 10^6$ bits.
This estimate is supported by the synthetic minimal cell JCVI-syn3.0, developed from {\it Mycoplasma mycoides}, which contains only 473 genes \cite{Hutchison2016}. Remarkably, 149 of these genes still have unknown function—underscoring how much we have yet to understand, even in minimal life.

To estimate $I_s$, we consider protein structure and folding. A protein's biological function depends on its shape. While AlphaFold2—an AI-based protein-folding algorithm—occupies 556 GB, this includes thousands of structures, redundancies, and metadata \cite{Jumper2021}. A more realistic lower bound for a minimal cell might be in the range of $10^6$–$10^8$ bits, depending on the number of unique folds, protein domains, and the information needed to describe each.

To estimate $I_d$, the dynamic component, we turn to whole-cell models. The first comprehensive whole-cell simulation of {\it Mycoplasma genitalium} integrated 28 submodels covering DNA replication, transcription, translation, metabolism, and more \cite{Karr2012}. The compressed code size was about 140 MB, or roughly $10^9$ bits. The model used over 1,900 parameters curated from 900+ publications and could predict phenotypes from genotypes. While this is almost certainly an overestimate—due to redundant code and parameters—it's informative. Notably, the model excludes spatial organization, mechanical feedback, and memory effects. Other whole-cell simulations of {\it E. coli} \cite{AhnHorst2022} and synthetic cell JCVI-syn3A \cite{Domenzain2022} use somewhat smaller codebases but follow similar principles.

 {We now turn to $I_a$, the assembly component of the information required to make a protocell, and ask how unlikely it is for such an object to arise through random chemical assembly alone. Instead of invoking more elaborate assembly theories, we use a simple branching-process picture. We consider an object of effective assembly length $L$, with a branching factor $b$ describing the number of random chemical choices available at each step. The total number of possible assembly paths then scales as $N_{\text{paths}}\sim b^L$, while the number of paths that successfully produce the target object is small, $N_{\text{object}}$, plausibly scaling only polynomially with $L$, $N_{\text{object}}\sim \text{poly}(L)$, or even $\mathcal{O}(1)$ as a first approximation.}

 {The probability that a random assembly trajectory of length $L$ reaches the target structure is therefore
\begin{equation}
P_{\text{path}}(X)\sim \frac{L^K}{b^L}\approx b^{-L},
\end{equation}
up to polynomial prefactors. This expression formalizes the intuition that for complex objects of large size $L$, the set of successful assembly histories forms a vanishingly thin corridor within an exponentially large tree of possibilities. Taking the negative logarithm of this probability yields a natural estimate of the assembly information cost associated with realizing the target structure. This argument can be extended to $M$ modules assembled hierarchically.\footnote{For a complex structure composed of $M$ modules, one may write
$P_{\text{path}}(X)\sim \sum_i^M L_i^K/b_i^{L_i} + (\text{joining terms})$ for modules of size $L_i$ and branching factor $b_i$, with $i=1,\ldots,M$.}}

 {Plugging in some numbers for an order of magnitude estimate, we may say there are $M\sim 10^3$ modules, each typically made from $L\sim 10^3$ key effective steps, and an effective branching factor of $b_\text{eff}\sim 10-10^3$, describing possible reaction or attachments in the prebiotic chemical environment. Neglecting joining issues from the different modules for simplicity, this would lead to 
$I_a\sim M L \log_2(b_\text{eff})\sim 10^6 \times (3-10)$ bits $\sim 10^7$ bits, but, due to the immense difficulty in estimating this, could equally lead to larger numbers like $10^9$ bits similar to $I_s$ or even larger, forming the bottleneck of protocell formation. Hence, a cell is not only informationally complex in its final configuration; it also sits at the end of an astonishingly long and narrow assembly corridor in process space.}

Taken together, we estimate the total information content of a protocell to be $I_\text{cell} \sim 10^9$ bits. This yields a minimum required information accumulation rate of
\begin{equation}
R_{\text{min}} = \frac{I_{\text{protocell}}}{T_\text{available}} \approx 6.34 \times 10^{-8} \ \text{bit/s},
\end{equation}
corresponding to an accumulation rate of only about 2 bits per year.
Compared to $R(D) \sim 100$ bit/s (modulo the efficiency factor $\eta$), this is minuscule—but reflects the vast timescale over which life could have emerged. To illustrate: a DNA molecule grows by 2 bits per base pair. At a rate of 100 bit/s, this corresponds to 50 base pairs per second, or more than the considerable $500$ m length of a DNA molecule in a year. The remarkably low threshold for information accumulation at $6.34 \times 10^{-8}$ bit/s suggests that abiotic emergence of life should be feasible, even with extremely low efficiency ($\eta \sim 10^{-8}$) or short molecular lifetimes ($D \sim 1$ s).

\section*{Drift versus random diffusive process for self-assembly}

\noindent The minimum information accumulation rate required to assemble a protocell of complexity $I_{\text{protocell}} = 10^9$ bits within $T_\text{available} = 500$ million years is
$v_{\text{min}} = R_{\text{min}} = 6.34 \times 10^{-8}$ bit/s,
highlighting that even extremely inefficient or sporadic processes could, in principle, accumulate the required information over geological timescales. However, this estimate assumes a unidirectional, steadily progressive (ballistic or forward-biased) process. What if the information-gathering process behaves more like a random walk—with fluctuations and reversals—rather than directed drift?

\begin{figure}[t]
\centering
\includegraphics[scale=0.51]{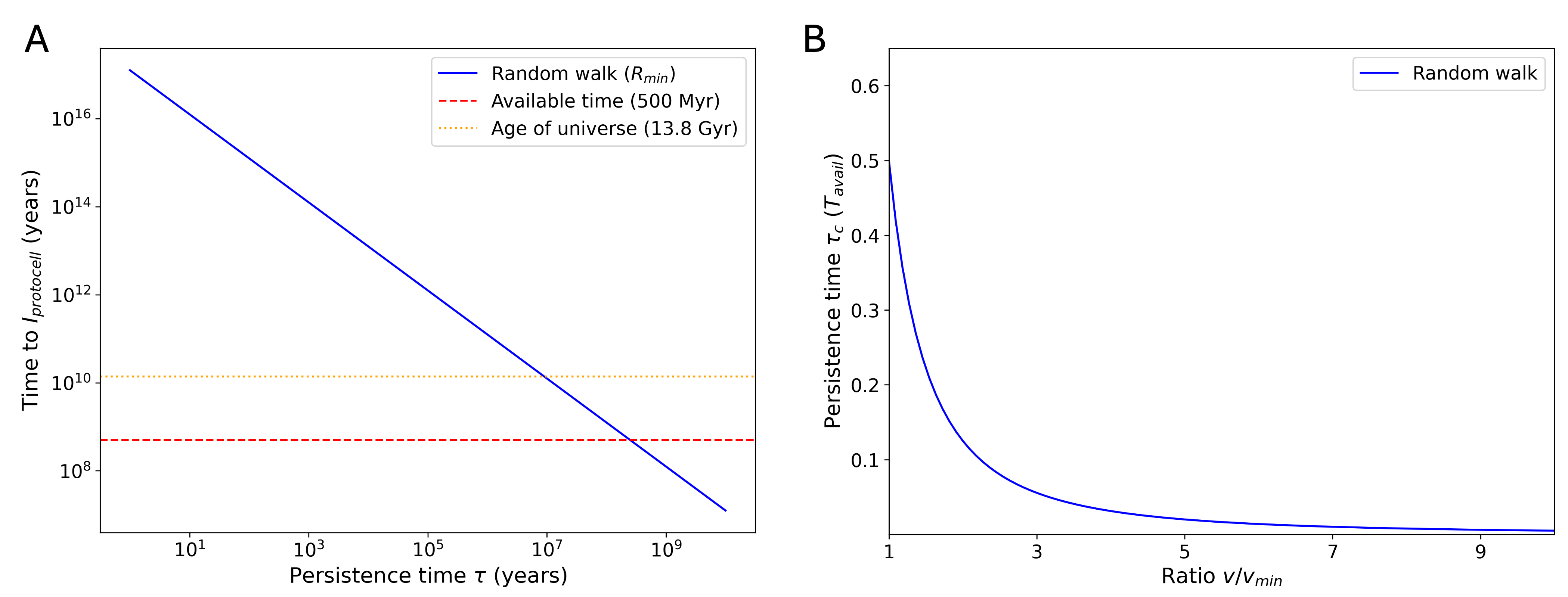}
\caption{{\bf How persistent does a random walk need to be in information space?} (A) Time to reach protocell complexity $I_\text{protocell}$ as a function of the persistence time $\tau$ for a random walk using minimal information accumulation speed  $R_\text{min}$, given by $T' = \langle I_{\text{protocell}}^2\rangle/(2 v_{\text{min}}^2\, \tau)$ (blue line). The available time of $500$ Myr (red dashed line) is reached for a persistence time of $250$ Myr. Also shown is the age of the universe (yellow line).  {(B) Persistence time for allowing larger information accumulation speeds than $v_\text{min}$ for protocell formation in $T_\text{available}$. The larger the speed of information accumulation, the shorter the required persistence time of the random walk. For details see the github repository.}}
\label{fig:fig3}
\end{figure}

Let us consider a stochastic accumulation model: a random walk with velocity $v$ and persistence time $\tau$, analogous to “run-and-tumble” dynamics from bacterial chemotaxis \cite{Berg1993}. The effective diffusion constant is given by $D_{\text{eff}} = v^2 \tau/d$,
where $d$ is the dimensionality of the search space. If $v = v_{\text{min}}$, then the mean-square displacement (MSD) after time $T_\text{available}$ is
\begin{equation}
\langle I^2_{\text{protocell}} \rangle = (v_{\text{min}} T_\text{available})^2 = 2 d D_{\text{eff}} T_\text{available}  = 2 v_{\text{min}}^2 \tau T_\text{available},\label{eq:tau_c}
\end{equation}
independent of $d$. Solving for the crossover time $\tau_c$ at which this diffusive process accumulates the same information as directed drift yields
$\tau_c = T_\text{available}/2$. Thus, if $\tau < \tau_c = 250$ Myr, the randomization is too strong, and the process fails to reach the target within $T_\text{available}$ (see Fig. \ref{fig:fig3}A). High persistence—i.e., strong directional memory—is essential. To illustrate the timescale sensitivity: With $\tau = 1$ s, the wait time for assembling a cell becomes $10^{24}$ yr — roughly a hundred trillion universes stacked end to end. Even if $\tau = 1$ yr, the required time is still $T' \approx 10^{17}$ yr, about ten million times the universe’s current age. In other words, without immense persistence, life’s emergence becomes cosmologically implausible, potentially pointing to alternative mechanisms.

 {However, this result may be too restrictive as we solely allowed minimum information accumulation speed $v_\text{min}$ which is determined to make the drift scenario just work in the available time $T_\text{available}$. A persistence time of $\tau_c=T_\text{available}/2$ simply reflects that, after doing one initial step towards a protocell, the next step, another gigantic "fluctuation" of random organization over $T_\text{available}/2$, either completes the protocell, or undoes any progress made, leading to a completed protocell with 50\% chance. Let's allow $v>v_\text{min}$ and hence $\tau_c'<\tau_c$ via Eq. \ref{eq:tau_c}  with $v_\text{min}$ replaced by $v$ on the right-hand side and solving for $\tau=\tau_c'$. Now, $\tau_c'=(v_\text{min}/v)^2 T_\text{available}/2$. The result is shown in Fig. \ref{fig:fig3}B. Making a protocell in $T_\text{available}$ with a persistence time of e.g. "only" $\tau_c'=1,000$ yr now requires $v=500\times v_\text{min}\approx 3.2\cdot 10^{-5}$ bit/s (or equivalently of about $500$ bp/yr), and for $\tau_c'=1$ s (still large for molecular reaction time scales) now requires $v\approx 6$ bit/s (or $2.8$ bp/s). Hence, while some of those persistent times and speeds seem reasonable to make a protocell by random walk, this has to be done consistently for $500$ Gyr, even though the climate is fluctuating, especially on the early planet Earth.}

 {Up to now we laid out the main contribution of this paper: the emergence of a protocell is possible, but only if information accumulation via random walks has a large persistence time (or a high effective information accumulation rate), over extended (enormous) periods of time. However, the devil is in the details, specifically factor $\eta$ - the actual physio-chemical mechanism and the conceptual big unknown. This factor could be close to zero, rendering the rate–distortion bound unattainable in practice. In the following we examine several factors influencing $\eta$, including energetic requirements (which appear comparatively permissive), assembly kinetics (which likely constitute the dominant bottleneck), and nonequilibrium phase transitions (which may offer a potential route to overcoming these limitations).}

\section*{Material uptake and energy consumption are not limiting}

\noindent To assess whether the prebiotic environment could support the molecular uptake rate required by our information-theoretic model, we compare it to Berg and Purcell's classical  estimate for the molecular flux to a spherical cell \cite{BergPurcell1977}. For a cell of radius $a = 1 \, \mu\mathrm{m}$, diffusion coefficient $D = 300 \, \mu\mathrm{m}^2/\mathrm{s}$, and concentration $c = 1 \, \mathrm{mM}$, the diffusive flux is given by $J = 4 \pi D a c \approx 2.27 \times 10^9 \, \text{molecules/s}$. 

In contrast, our rate-distortion estimate requires an information accumulation rate of $R(D) \sim 100 \, \text{bit/s}$, with each molecule contributing approximately 10 bits. This implies a necessary uptake of only $R(D)/10 = 10$ molecules/s. The actual molecular flux thus exceeds the required rate by a factor of approximately $10^8$. Therefore, the prebiotic chemical environment was not transport-limited—even at nanomolar concentrations. Rather, the challenge lay in filtering, stabilizing, and retaining the few informationally useful molecules from the overwhelming molecular background.

Using the minimum required information rate $v_{\text{min}} \approx 6.34 \times 10^{-8}$ bit/s and again assuming 10 bits per useful molecule, the corresponding capture rate is $v_\text{min}/10 \approx 6.34 \times 10^{-9}$ molecules/s. This implies that, on average, fewer than one molecule needs to be captured every 5 years. Compared to the Berg and Purcell limit of $10^9$ molecules/s, this rate is 13 orders of magnitude lower. Hence, even extremely sparse and selective molecular capture would suffice to meet the informational demands of life's emergence, given enough time.

If a protocell requires $I_{\text{protocell}} \sim 10^9$ bits, and each molecule contributes approximately 10 bits, then the total number of molecules needed is $N_{\text{protocell}} = I_{\text{protocell}}/10 = 10^8$ molecules, consistent with estimates for {\it E. coli} \cite{Milo2015CellBiologyByNumbers}. Assuming diffusion-limited uptake, the time required to accumulate $N_{\text{protocell}}$ molecules is $\Delta t = N_{\text{protocell}}/J = 10^8/(2.27 \times 10^9) \approx 0.044 \ \text{s} \approx 1.4 \times 10^{-9}$ yr. In principle, a protocell could acquire all necessary components in under 50 milliseconds—if every incoming molecule were retained and useful. In reality, however, most molecules would be non-functional, degraded, or unincorporated, making molecular selection and retention—not flux—the true bottlenecks in prebiotic information accumulation.

To estimate the energy required to  {make} a system with $I_{\text{protocell}} \sim 10^9$ bits, we compare three frameworks: the Landauer limit, biological ATP-driven synthesis, and technological processes such as PCR. First, the Landauer principle states that erasing one bit of information incurs a minimum energy cost of $E_{\text{min}} = k_B T \ln 2 \approx 2.8 \times 10^{-21}$ J/bit at $T \approx 300$ K. Thus, the theoretical minimum energy for  {making} the protocell is only $E_{\text{Landauer}} \approx 10^9 \times 2.8 \times 10^{-21} = 2.8 \times 10^{-12} \ \text{J}$. Second, biological energy costs based on ATP hydrolysis provide about $10^{-19}$ J per molecule. If synthesizing one informational molecule (containing $\sim$10 bits) requires one ATP (a very conservative assumption), then $E_{\text{ATP}} = 10^8 \times 10^{-19} = 10^{-11} \ \text{J}$. Third, technological synthesis via polymerase chain reaction (PCR) consumes about 1–10 J to amplify microgram-scale DNA. A 10$^9$-bit construct corresponds to $\sim$50 $\mu$g of DNA, implying $E_{\text{PCR}} \sim 1 \ \text{J}$, or roughly $10^{-9}$ J/bit. Even 1 J is a small amount of energy—equivalent to lifting a 100 g apple by 1 meter or powering a smartphone for 1 second.

 {As a formal aside, and solely to obtain a mechanistic lower bound on energetic cost, it is useful to frame the problem in terms of unbalanced optimal transport \cite{UOT2016}. Let the environment contain molecular building-blocks of types $i=1,\dots,B$ with abundances $\mu_i$, and let a protocell require molecular species $j=1,\dots,N$ in quantities $\nu_j$. A transport--transformation plan $\pi_{ij}$ specifies how many molecules of type $i$ are converted into molecules of type $j$. The unbalanced case is appropriate here because the environment generally does not contain enough of the specific molecules needed; creation of new molecules from simpler precursors is therefore required. A simple unbalanced optimal transport cost functional is
\begin{equation}
\mathrm{cost}(\pi)
=
\sum_{i,j} c_{ij}\,\pi_{ij}
\;+\;
\kappa_{\mathrm{tgt}}\sum_{j}\left(\nu_j - \sum_{i}\pi_{ij}\right)_{+},
\label{eq:OTcost}
\end{equation}
where $c_{ij}$ denotes the free-energy cost of transforming a molecule of type $i$ into one of type $j$, and $\kappa_{\mathrm{tgt}}$ is the free energy required to synthesize a molecule of type $j$ ``from scratch'' if the environment does not supply enough precursors. The $(\cdot)_{+}$ denotes the positive part, ensuring that shortages must be paid for by {\it de novo} synthesis. This functional has a direct physical interpretation: it computes the minimal reversible Gibbs free energy required to convert an environmental composition vector $\mu$ into the target protocell composition $\nu$, assuming perfectly efficient chemistry, perfect selectivity, and no kinetic or spatial constraints.}

 {For present purposes it is useful to make the most conservative assumption possible, namely that the environment does not conveniently provide large amounts of completed lipids, nucleotides, or peptides. In this case virtually all material must be created by de novo synthesis, and the optimisation of~\eqref{eq:OTcost} collapses to a sum over required protocell components,
\begin{equation}
E_{\min}
\simeq
N_L\,\Delta G_L
+
N_N\,\Delta G_N
+
N_P\,\Delta G_P,
\label{eq:EminSimple}
\end{equation}
where $N_L$, $N_N$, and $N_P$ are the numbers of membrane lipids, nucleotide monomers, and amino-acid residues required, respectively. The quantities $\Delta G_L$, $\Delta G_N$, and $\Delta G_P$ represent lower-bound reversible free energies for synthesising each corresponding molecule from simple bulk precursors (CO$_2$, H$_2$, N$_2$, etc.). These free-energy values implicitly include the minimal chemical work, the minimal transport work (such as insertion of lipids into a membrane), and the minimal assembly work (such as forming a phosphodiester or peptide bond), and are deliberately chosen to be small so that the resulting estimate is an extremely generous lower bound.}

 {To obtain a numerical value, consider a toy protocell containing $N_L \sim 10^8$ lipids, $N_N \sim 10^6$ nucleotide monomers, and $N_P \sim 10^6$ amino-acid residues. Conservatively taking $\Delta G_L \sim 50\,\mathrm{kJ/mol}$, $\Delta G_N \sim 60\,\mathrm{kJ/mol}$, and $\Delta G_P \sim 40\,\mathrm{kJ/mol}$, and converting from molar to per-molecule scales, one finds
\begin{equation}
E_{\min} \sim 10^{-11}\,\mathrm{J},
\end{equation}
which is already the most favourable reversible estimate obtainable from any physically meaningful choice of parameters. Expressed in thermal units at $T\approx 300\,\mathrm{K}$,
\begin{equation}
\frac{E_{\min}}{k_B T} \sim 10^{9},
\end{equation}
and using the Landauer relation $k_B T\ln 2$ per bit, this corresponds to an effective thermodynamic complexity of approximately
\begin{equation}
\frac{E_{\min}}{k_B T\ln 2} \sim 10^{9}\ \text{bits}.
\end{equation}}

 {It is important to emphasise that this $E_{\min}$ represents only a lower bound on the minimal reversible work of assembly. It neglects all kinetic barriers, all dissipation, all side reactions, all issues of pathway search, all spatial transport limitations, all problems of chirality purification, and all requirements for templating or sequence order. Any realistic prebiotic scenario would require substantially more energy, but this most optimistic bound lies in the range suggested by independent arguments such as energy requirements ranging from $10^{-12}-10^{-9}$ J (or even $1$ J for modern inefficient PCR) and $I_a\sim 10^7-10^9$ bits.
Hence, we are beginning to paint a consistent picture for evaluating the feasibility of prebiotic assembly.}

\section*{Finding the assembly path is the tricky bit}

 {Having argued that neither material flux nor energetic cost poses a fundamental obstacle, we now turn to what remains: the structure of assembly paths themselves. Earlier we used a simple branching process to estimate the complexity involved in assembling a protocell and then translated this into an energetic cost using the Landauer bound, yielding thermodynamic complexity estimates that are, in principle, manageable and consistent with earlier bounds. In reality, however, the assembly process is far more subtle. Molecules must form complex structures through hierarchical assembly in a manner that does not obstruct subsequent steps, and the entire process must function dynamically—as a machine in motion rather than a frozen sculpture.}

 {To gain basic intuition, this assembly process is rather very different from assembling a 2D jigsaw puzzle. Even if a piece is missing in the middle, we can easily reach that space and fill it using the available third dimension; we can simply move it there, not affecting the remainder of the puzzle. The assembly path is simple and highly parallel: we can start in any corner, or anywhere in the middle, or do it in chunks and assemble the chunks later.}

 {To formalize this difficult process, we turn to empirical assembly theory \cite{Sharma2023Assembly}\footnote{While a useful framework, 'theory' is a strong word - it is more like an empirical metric of complexity.}, which is a framework that quantifies complexity by measuring the minimal number of assembly steps needed to build an object from its basic parts. It suggests that complex objects, especially those with biological origins, are unlikely to arise by random chance and require a specific, historical path to be built. The framework provides a measurable way to distinguish between objects made by random processes and those created through selection and evolution, and it has implications for detecting life on other planets.}

 {In this framework, the assembly number $A$, is an empirical measure designed to quantify how strongly a molecule is supported by selection and replication versus passive abiotic chemistry \cite{Sharma2023Assembly}. For a given molecular species $i$ with assembly index $a_i$ (the length of the shortest construction path in a restricted combinatorial space) and abundance $n_i$, they define
\begin{equation}
A \;=\; \sum_{i=1}^{N} \exp(a_i)\,\frac{n_i - 1}{N_T},
\label{eq:Adef}
\end{equation}
where $N_T=\sum_in_i$ is the total number of molecules detectable above threshold. Although the expression resembles a Boltzmann weight, the factor $\exp(a_i)$ is not thermodynamic; it is an empirical amplifier that gives exponentially greater weight to complex molecules (large $a_i$), while the factor $(n_i - 1)/N_T$ encodes relative abundance. Thus $A$ is best viewed as a ``complexity $\times$ abundance'' score: large $A$ values indicate complex molecules that are also unexpectedly common relative to abiotic expectations, whereas small $A$ values correspond either too simple or too complex molecules whose abundance is negligible.}

 {This structure predicts a characteristic bimodality when life is present. Abiotic chemistry alone is efficient at generating very low-assembly-index structures, but strongly suppresses high-assembly-index molecules. Under selection and replication, however, specific high-$a$ structures become amplified, creating a second peak in abundance. Between these two peaks, one generically expects a region of extremely low abundance---an ``assembly desert''---because there is no smooth path of intermediate molecules that are both synthesized frequently enough and sufficiently stable. In other words, abiotic chemistry efficiently populates $a\approx 1$--$4$; biology populates $a\gtrsim 20$; and the range in between is nearly empty.}

 {In a simple toy model of only one key molecular species, Eq. \ref{eq:Adef} simplifies to $A\sim\exp(a)\;n$. This picture aligns with a simple branching-path model from earlier in which the probability of generating a molecule of assembly index $a$ by purely random, unguided exploration scales as
$n_{\text{abiotic}}(a)\;\propto\;P_{\text{abiotic}}(a)\;\propto\; b^{-a} \,\mathrm{poly}(a)$,
where $b$ is the average branching factor of the assembly graph, combined with a mild combinatorial prefactor of polynomial in $a$. 
The assembly index for possible structures of size $a$, we obtain
\begin{equation}
A_{\text{abiotic}}(a)\;\sim\;
\left( \frac{e}{b} \right)^{a}\,\mathrm{poly}(a).
\label{eq:Aabiotic}
\end{equation}
Since $b \gg e$ for any realistic molecular construction space, the factor $(e/b)^{a}$ decreases super-exponentially, so $A_{\text{abiotic}}(a)$ is sharply concentrated at very low $a$ and becomes effectively zero for modest $a\gtrsim 10$. This mathematically explains the expected emptiness (the ``desert'') in the intermediate region.}

\begin{figure}[t]
\centering
\includegraphics[scale=0.55]{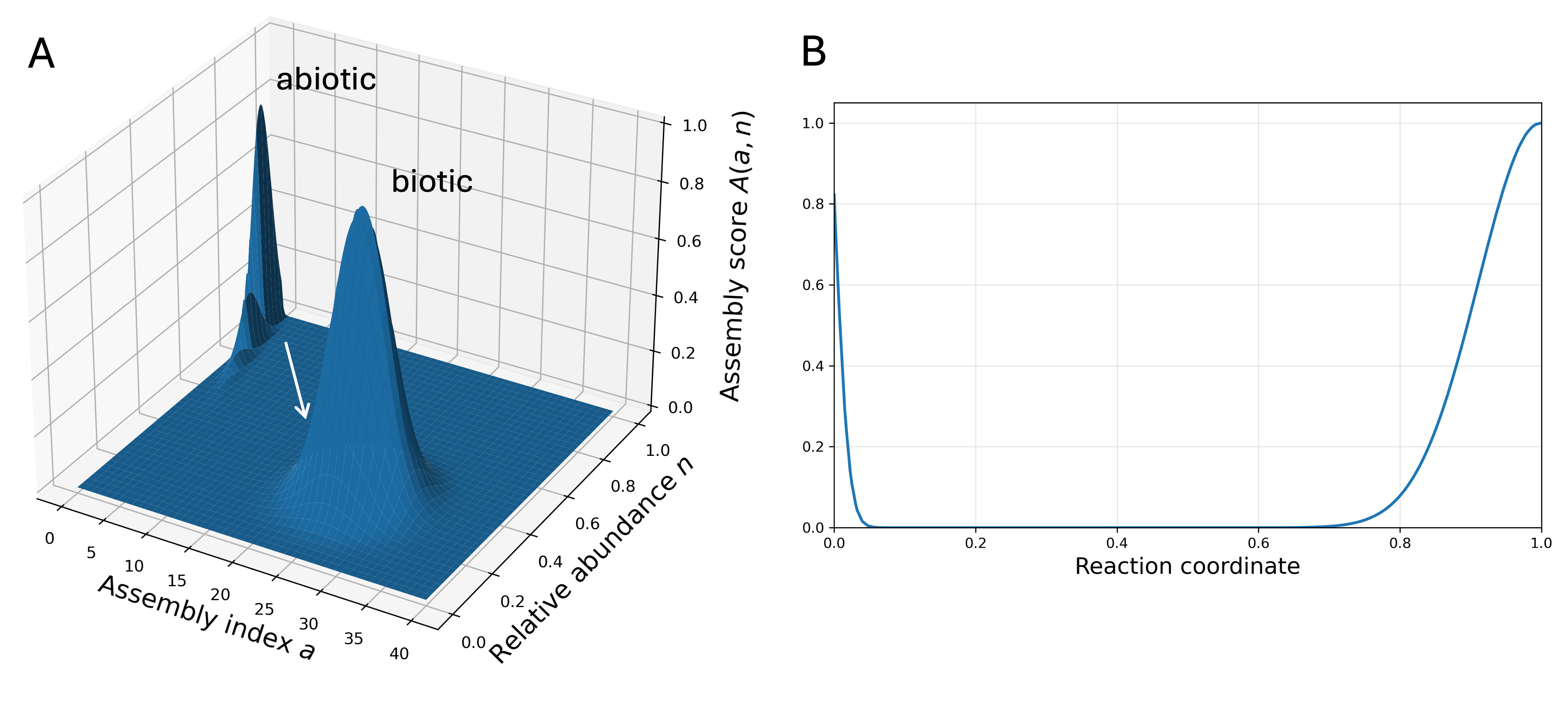}
\caption{ {{\bf Assembly theory illustrates the immense difficulty of assembling a protocell.} (A) Assembly score from a toy model as a function of assembly index $a$ and relative abundance (copy number) $n$, for a single type of molecule (e.g. representing a bottleneck in assembly). There is a peak at low $a$ and high $n$ for abiotic chemistry, and another peak at high $a$ and intermediate $n$, representing a signature of life. To obtain such $a$ and $n$ values requires natural selection and growth. The white arrow indicates a kind of reaction coordinate connecting the two peaks. (B) Assembly score plotted along the reaction coordinate, illustrating the vast gap between the peaks. Mathematical details:  plot shows  
$A(a,n)\propto e^{\alpha a}\;n\;(\rho_{\mathrm{abiotic}}+\rho_{\mathrm{life}})$, 
normalized to its maximum, based on an abiotic density 
$\rho_{\mathrm{abiotic}}(a,n)=b^{-a}\exp[-(n-n_0 e^{-\gamma a})^2/(2\sigma_n^2)]$ 
and a biotic Gaussian bump at high $a$ and moderate $n$, for suitable parameters $\alpha$, $b$, $n_0$, $\gamma$, $\sigma_n$ etc. For details see the github repository.}}
\label{fig:fig4}
\end{figure}

 {In contrast, biology does not obey the branching process distribution. Once self-replication begins, abundances are driven by selection, not by combinatorics. For a molecular species replicated during metabolism or encoded by genes,
$n_{\text{biotic}}(a)>\!\!>n_{\text{abiotic}}(a)$. Even species with very high assembly index may have many copies, produced repeatedly, stabilized, and protected from degradation. Plug that into the assembly index and we obtain:
\begin{equation}
A_{\text{biotic}}(a)\;\sim\;e^{a}n_{\text{biotic}}(a).
\label{eq:biotic}
\end{equation}
Even moderate amplification of complex molecules causes a huge increase in $A$ because $\exp(a)$ is already exponential. Combined with even a modest abundance $n_{\text{biotic}}(a)$, a second peak emerges. This is the “life signature.” Figure \ref{fig:fig4} shows results from a toy model, illustrating the two peaks for a single molecular species with assembly index $a$ and relative abundance $n$ (see caption for mathematical details).}

 {The existence of such an empty bridging region is important for the origin-of-life problem. If there were a smooth, gradually decreasing abundance curve from low to high assembly index, then high-$a$ biomolecules could plausibly arise from incremental abiotic processes. Instead, both theory and observation suggest a discontinuity: a sparsely populated or empty interval separating simple molecules from the complex macromolecules that biology produces in abundance. This gap indicates that unguided prebiotic chemistry cannot efficiently traverse the exponentially expanding assembly space, and that only selection, templating, and replication can maintain a stable population of large-$a$ structures. The ``assembly desert'' therefore acts as a signature of life and simultaneously marks a fundamental barrier that abiotic chemistry must cross for life to emerge.}

 {Unlike gaps in the fossil record, which can be explained by taphonomic biases and punctuated evolutionary equilibria \cite{GouldEldredge1972Punctuated, EldredgeGould1997PEAge, BentonHarper2009Paleobiology, Prothero2007Fossils, Raup1976SpeciesDiversity}\footnote{Speciation events require small, isolated populations, but fossilization requires large, stable, sediment-rich environments — the two rarely coincide.}, the absence of intermediate-complexity molecules in modern environments is a genuine chemical signal. Molecules do not fossilize, and if abiotic chemistry readily produced mid-complexity polymers, co-factors, nucleic-acid analogues, or protocell precursors, they should still occur today in hydrothermal vents, volcanic ponds, atmospheric photochemistry, or mineral interfaces \cite{Benner2010DefiningLife,Sutherland2017EndBeginning, KitadaiMaruyama2018OriginsBlocks}\footnote{Unless early Earth was very different from today's extreme conditions, or complex molecules are constantly generated but we don't look at the right place (e.g. environmental protective niches which can act as scaffolds of assembly), or we don't recognize such complex molecule-breeding grounds. Most likely, however, the common sense answer is No - chemical complexity of intermediate level is not occurring anywhere at a high enough rate.} . Yet modern Earth shows a stark bimodality: abundant simple molecules produced by geochemistry, and abundant complex molecules produced by life, with almost nothing in the middle \cite{Lambert2008PrebioticPolymerization,DangerPlassonPascal2012Peptides, Cleaves2008PrebioticAminoAcids, 
PownerSutherland2008OnePotNucleotides,
Bernhardt2012RNAWorldFragile}. This “assembly desert” cannot plausibly be explained by biological consumption, because cells do not import or metabolize arbitrary exotic polymers (e.g., PNA, TNA, random peptides) \cite{LazcanoMiller1999Metabolism, KooninMartin2005OriginGenomes}. Rather, it reflects a genuine chemical bottleneck: the mid-complexity region is extremely difficult to populate without templating, compartmentalization, or sequence-controlled polymerization \cite{Hazen2007emergence, Sasselov2020PlanetaryPhenomenon}. The absence of such molecules today is therefore evidence that the barrier they present is fundamental, and that early life must have required special mechanisms to cross it \cite{CleavesBada2012AlternativeNucleicAcids}.}

 {Overall, assembly theory highlights a striking structural feature of chemical space: the deep gulf separating the simple, abundant molecules of abiotic chemistry from the complex, abundant molecules of biology. Its bimodal landscape makes clear that bridging this “assembly desert” requires mechanisms beyond trivial chemical exploration, sharpening—rather than resolving—the central puzzle of how life first crossed from one peak to the other. This reframes the origin of life not as a problem of energy or material supply, but as a problem of accessing and stabilizing rare assembly pathways.}

\section*{Can information accumulate suddenly?}

\noindent Above, we estimated the information content of a minimal protocell using Kolmogorov complexity with some considerations of assembly, based on detailed quasi-realistic descriptions of cellular processes encoded in modern computational models. This supports a picture of life emerging through a gradual, steady increase in the organization and complexity of matter—from nonliving to living. This is like putting together a  {gigantic jigsaw puzzle}, and appears like the culmination of a reductionistic belief system. And the required persistence times are extreme, even mind boggling.

An alternative possibility is that life arose via a rather sudden explosive-like transition — by some yet-unknown self-organization principle of collectiveness triumphing over reductionism, that bridged the gap between random chemistry and selective evolution, potentially circumventing the information bottleneck of early self-replication. This idea is not new. It is most well developed in the study of collectively autocatalytic sets, first introduced by Kauffman~(1986) \cite{kauffman1986autocatalytic}. In such networks, each molecule is catalyzed by another member of the set, enabling system-wide replication without requiring any single molecule to self-replicate or the presence of a membrane-enclosed protocell. This framework offers a resolution to Eigen’s paradox: long, information-rich molecules require error correction to persist, but error-correcting machinery itself demands such complexity \cite{eigen1971self}. Autocatalytic sets enable increasing functional complexity before templated replication emerges. Mathematical models show that as chemical diversity increases, the probability of forming an autocatalytic set undergoes a sharp phase transition—a sudden onset of self-sustaining autocatalysis \cite{hordijk2004detecting,filisetti2011stochastic}. This is analogous to a percolation transition in random graphs.

Such transitions may also explain later leaps in complexity, including the emergence of the brain. Put simply, a brain functions as an internal model of the external world, trained by experience and used to predict future outcomes \cite{hidalgo2014information}\footnote{This predictive capacity has been formalized in the Bayesian brain hypothesis \cite{Knill2004BayesianBrain}, where internal priors are continuously updated in light of new data, approximating posterior beliefs that guide perception and action.}. For example, one can model parameter adaptation in a toy brain using the inverse temperature $\beta$ in an Ising-like system, rather than connectivity or size, as the critical control parameter. In such models, evolutionary fitness peaks near a critical value of $\beta$ when environments are dynamic, heterogeneous, and unpredictable \cite{hidalgo2014information}. Though phase transitions technically apply to infinite systems, the benefits of sharp transitions often hold in large finite systems as well \cite{hidalgo2014information,depalo2017critical,mora2011criticality}.

\vspace{1ex}
In the spirit of this article, we now return to AI. Suppose that chemical reaction networks in a cell can be viewed as a kind of neural network. Then we may ask: how much information must be encoded in such a network to achieve life-like complexity? Recent work on recurrent neural chemical reaction networks (RNCRNs) has shown that artificial recurrent neural networks can be mapped onto chemical reaction networks \cite{dack2024rncrn}. These networks are capable of approximating arbitrary dynamical systems governed by ordinary differential equations. Unlike feedforward neural networks, which approximate static input–output mappings, RNCRNs model time-dependent behaviors such as switching in multi-stable systems or oscillations frequently observed in biology.

This suggests that sufficiently complex chemical reaction networks can mimic neural networks—and hence approximate nearly any behavior, from fast signaling to slow gene regulation and cell-cycle control. This relies on the universal approximation theorem, which guarantees that a neural network with sufficient size can approximate any continuous function to arbitrary accuracy. In this view, a collection of interacting molecules becomes a cell once it can approximate adaptive functions required for survival.

Currently, such RNCRNs are small ($<10$ species or nodes) \cite{dack2024rncrn}, and it remains unclear how large a network must be to compute the full repertoire of cellular behavior. Still, we can try to estimate the required network size. The universal approximation theorem states that for any continuous real-valued function and any target error $\varepsilon > 0$, there exists a neural network that approximates $f$ within $\varepsilon$ on a compact domain. Recent results give an upper bound on the number of neurons $M$ needed to achieve this approximation:
\begin{equation}\label{eq:Meps}
  M(\varepsilon) \leq C \cdot \varepsilon^{-n/k},
\end{equation}
where $n$ is the input dimension, $k$ is the smoothness of the function class, and $C$ is a constant depending on $f$. Smoother functions (higher $k$) require fewer neurons. However, as the input dimension $n$ increases, approximation becomes harder—the well-known curse of dimensionality.

How does this network size map to information content? If each parameter (e.g., weight or bias) must be specified with $p$ bits of precision, then the total information required to define the network is
\begin{equation}\label{eq:I_NN}
  I_\text{NN} \approx M \cdot p.
\end{equation}
Combining Eqs.~\ref{eq:Meps} and \ref{eq:I_NN}, we obtain:
\[
I_\text{protocell}(\varepsilon) \sim p \cdot \varepsilon^{-n/k}.
\]
This links approximation error $\varepsilon$ directly to informational complexity, assuming uniform precision $p$. Note that this estimate applies only to the dynamic regulatory core of the protocell—not its full genomic, structural, or metabolic description. This framework offers a principled way to estimate the minimal informational content of life-like systems, constrained by approximation theory and encoding limits. When combined with rate-distortion theory, it may help formalize how much structure, memory, and robustness a protocell must have to be functionally complete. However, there is no prediction of a sudden transition to higher complexity.

\vspace{1ex}
Let us now take a different route and investigate how large and connected a network must be for computation to emerge spontaneously. A linear threshold model exhibits a percolation transition at a critical mean degree $z_c = 1$. This means that when, on average, each node is linked to one other, a giant connected component (GCC) for nontrivial computation appears  \cite{wilkerson2022spontaneous}. A minimal network of approximately $10^4$ components—if sufficiently connected and properly balanced between activation and inhibition—can begin to compute complex Boolean functions. This supports the idea that a protocell capable of decision-making may require only $10^4$ interacting components. If each contributes 10 bits, the total is just $10^5$–$10^6$ bits—well below the earlier estimate of $10^9$ bits. But here, we consider only logical capacity, not structural or biochemical implementation, making this a lower bound.

\vspace{1ex}
Hence, sufficiently complex chemical reaction networks can mimic the dynamics of recurrent neural networks, or make the emergence of complex logic functions inevitable. This raises the intriguing possibility that biological systems—even at the molecular scale—approach universal computation in their expressive power. Indeed, chemical kinetics, when properly structured, can implement universal computation \cite{magnasco1997chemical}. More concretely, chemical reaction networks governed by mass-action kinetics can simulate arbitrary computations and are thus Turing complete \cite{soloveichik2008computation}.

 {
Taken together, these various exotic phase transitions leave the door open for the spontaneous assembly of protocells far from equilibrium. As collective emergent phenomena, such transitions are, by definition, difficult—if not impossible—to predict from mechanistic ingredients and first principles. While this may offer a route around the assembly bottleneck, it also leaves the ultimate watershed moment of life’s origin clouded in complexity-science mystery.
}

\section*{Discussion}

Setting aside the statistical fluke argument in an infinite universe, we have explored the feasibility of protocell self-assembly on early Earth. A minimal protocell of complexity $I_\text{protocell} \sim 10^9$ bits could, in principle, emerge abiotically within Earth’s available timespan ($\sim 500$ Myr)—but only if a tiny fraction of prebiotic interactions ($\eta \sim 10^{-8}$) are persistently retained over vast stretches of time. We approached this question through a conceptual framework combining rate-distortion theory, chemical entropy estimates, and Kolmogorov complexity derived from whole-cell and AI models.

To better understand what must be true for abiotic life to succeed, we proposed a ``requirements checklist'': (i) some degree of physical or chemical bias (e.g., compartments, cycles, autocatalytic networks), (ii) sufficient persistence time if information accumulates via a random walk through chemical space, and (iii) protection and reuse of functional molecules. If steady progression is too slow, a phase transition may have enabled a more abrupt leap in complexity. Either way, a purely random soup is too lossy—some form of prebiotic informational structure must precede Darwinian evolution.

Naturally, the framework has limitations. Our estimates—of entropy, distortion time, and protocell complexity—are crude at best, drawn from analogies rather than direct measurement. The entropy of the prebiotic chemical pool ($H_\text{prebiotic}$) is inferred from analogues like meteorites and Titan’s atmosphere, and the estimated information content of a minimal cell ($I_\text{protocell}$) is based on compressed models rather than any physical lower bound.  {In particular, the value of factor $\eta$, likely tiny, represents a catch-all for the yet-unknown mechanism.} Still, the exercise offers a quantitative scaffold for framing the problem—and perhaps guiding future investigations.

Beyond parameter uncertainty, as mentioned, the deeper challenge is mechanistic: even if the information rate is feasible, the route remains opaque. Where did the directionality (drift $v$ or persistence $\tau$) come from? What structures or environmental constraints enabled long-term memory or error suppression without evolved proofreading? The puzzle deepens as timelines shift: from 3.465 Gy microfossils in Western Australia \cite{Schopf2006,Schopf2007,Schopf2018} to a LUCA possibly living around 4.2 Gy ago \cite{Moody2024}, close to the formation of liquid water at 4.404 Gy \cite{Wilde2001}. But LUCA is just a branching point. Substantial evolutionary development—possibly involving other, now-extinct lineages—likely occurred between the first protocell and LUCA.

Which returns us, cautiously but irresistibly, to the question: Was Earth terraformed, or did order coalesce from chaos under the silent governance of physics? Today, humans seriously contemplate terraforming Mars or Venus in scientific journals \cite{sagan1961planetary,sole2020synthetic}. If advanced civilizations exist, it is not implausible they might attempt similar interventions—out of curiosity, necessity, or design. Still, Occam’s razor weighs in: abiotic evolution, however slow and strange, remains a viable (if mind-bending) explanation. Invoking terraforming adds explanatory complexity without constraint. And while we cannot prove that abiogenesis is inevitable, it remains consistent with thermodynamics.

If life is the ultimate emergent phenomenon, perhaps it resists prediction not because the physics is wrong, but because the framework is incomplete. As Laughlin has argued, emergence defies reductionism not in defiance of physics, but in fulfillment of it \cite{Laughlin2000MiddleWay,Laughlin2005DifferentUniverse}. The reaction path may be immensely long, contingent, and distributed across scales—too delicate to reconstruct after four billion years. In such a landscape, quantitative feasibility may be the most we can achieve. But feasibility, even without full predictability, can still inform what is physically possible—and that, for now, is enough to keep the question alive.

 {Looking ahead, we advocate for more empirical, exploratory work in the spirit of Miller and Urey—to better constrain persistence $\tau$, exploration (or drift) $D$ (or $v$), and mechanistic uncertainty $\eta$.} We suggest focusing on mechanisms of memory and information retention in prebiotic systems. For instance, in complex and stochastic networks, do certain configurations converge to robust attractor dynamics in molecular or ecological state space? Might this make evolution more deterministic than traditionally thought, as hinted at by recent outside-of-the-box experiments \cite{chuang2019homeorhesis,floroni2025membraneless}?

AI may itself play a decisive role. Beyond modeling chemical systems, it may help reverse-engineer candidate pathways, identify attractor landscapes, or uncover deep statistical regularities that escape human intuition. If life is a form of physical computation, as some have proposed \cite{Lloyd2000}, then AI might soon help identify how natural chemistry can become computationally powerful enough to self-organize. 

 {Although current AI systems excel at synthesizing knowledge and exploring enormous hypothesis spaces, their creativity remains largely one of rearrangement rather than genuine conceptual invention. I am personally sceptical that deep scientific insight can be achieved without the qualities that shape human creativity — purpose, passion, embodiment, and an awareness of our own finiteness. These give humans a sense of what truly matters and drive the irrational persistence behind transformative ideas. For instance, a lone genius might have a breakthrough driven by deeply personal, often irrational motivations \footnote{History offers countless examples in which creative persistence was fueled by romantic, social, or symbolic incentives rather than rational optimization. That kind of urgency or drive is hard to explain to an AI.}. Still, future AI may either approximate such motivational structures or overcome their absence through overwhelming exploratory capacity. For now, however, the kind of abductive leaps required to explain life’s emergence remain uniquely human.} Whether or not AI ultimately solves the origin-of-life problem, it may nevertheless turn out to be the next Cohen's microscope (or telescope), just better \cite{cohen2004mathematics}. Certainly, it has begun to rewrite the language in which biological complexity can be described.

 {
As a thought experiment, one can ask what a contemporary large language model might propose if tasked with addressing the origin-of-life problem under effectively unlimited resources over a decade. The model proposes a surprisingly coherent strategy. First, build a digital twin of early Earth—and of selected exoplanets—to assess whether our planet’s conditions were in any sense special. Second, create a vast “chemistry collider” capable of exploring prebiotic reactions at scale, in order to map the chemical landscape and understand the striking absence of mid-complexity abiotic molecules today. Third, systematically construct minimal forms of artificial life to determine what is truly required for sustained heredity, metabolism, and evolvability. Finally, establish a serious theory institute: a place where a few deeply creative individuals (“lone geniuses”), largely unconstrained by the burdens of everyday academic life, can think deeply about the problem, guided by the data above, and develop theoretical frameworks that go beyond purely brute-force approaches.
}

 {The striking point is that neither brute-force exploration nor isolated genius is likely to solve the problem on its own. Progress will almost certainly emerge from the interplay—perhaps the tension (likely frustration) — between large-scale big data with even bigger computers and small-scale conceptual innovation. Even so, we may never recover the precise historical sequence by which life first emerged. What we can hope for, however, is a set of physically and chemically plausible routes: an understanding of how matter of this kind could have become alive, even if we never know exactly how it did.}

We end with a note of caution. There is a real possibility that, in seeking to understand life’s origin, we become a living parable of Gödel’s incompleteness and Turing’s undecidability—systems entangled in their own logic, unable to fully explain themselves. Perhaps life cannot prove its own existence; perhaps the AI we task with the job will run forever, or halt with an answer as cryptic as ``42.'' To avoid standing dumbstruck—like an ape before a lightning-struck fire—we must ensure that our tools, however powerful, can still speak in terms we understand. Otherwise, we risk becoming spectators of intelligence we cannot follow, let alone guide.

\begin{quote}
\textit{``The miracle of the appropriateness of the language of mathematics for the formulation of the laws of physics is a wonderful gift which we neither understand nor deserve.''} — Eugene P. Wigner \cite{Wigner1960}
\end{quote}

\subsection*{Data and Code Availability}
 {No experimental data was produced in this work. All original code can be found in the GitHub repository https://github.com/Endres-group/origin-of-life-paper and is publicly available.}

\subsection*{Acknowledgements} 
We thank the Physics of Life Network of Excellence at Imperial College, for its stimulating seminars, journal club, and workshops. The latter included "Physics of emergent behaviour III: from origin of life to multicellularity" in 2021, among others.  {We also appreciate the hospitality at the Nature conference "AI augmented Biology" in Nanjing, China, in October 2025, which led to stimulating discussions.}
We also acknowledge the lack of funding, which, while generally detrimental, created space for pondering the fundamental questions of this article. 

\subsection*{Summary of changes in Version 2}
 {
Version~2 expands the conceptual scope of the paper by discussing the role of the global magnetic field, clarifying different photosynthetic pathways and their implications for the lifetime of the biosphere, and using the homochirality problem as an illustrative contribution to the efficiency factor $\eta$. It extends the protocell information estimate to include assembly-path complexity, refines the analysis of persistence times in random-walk-based assembly (with new Fig.~2B), introduces an unbalanced optimal transport bound on the minimal energetic cost of protocell formation, and adds a new section on assembly theory supported by a toy model (Fig.~3). The Discussion has been broadened to include a more critical perspective on the potential and limitations of artificial intelligence in origin-of-life research. Finally, a GitHub repository has been added and several references have been corrected.
}

\bibliography{references}

\end{document}